\documentstyle[12pt]{article}
\oddsidemargin--0mm
\begin{document}

\begin{center}
{\LARGE \bf Supersymmetry and Bogomol'nyi \\ equations  in the Maxwell
Chern-Simons systems}
\end{center}
\vskip 1cm
\begin{center}
 Bogdan Damski\footnote{E-mail address: bodzio@druid.if.uj.edu.pl}\\
\end{center}
\vskip 1cm
\begin{abstract}
 We take advantage of the superspace formalism and explicitly
 find the N=2 supersymmetric extension of the Maxwell Chern-Simons
 model.
In our construction a special form of a potential term and indispensability
  of an additional neutral scalar field arise naturally.
   By considering the algebra of supersymmetric
 charges we find Bogomol'nyi equations for the investigated model.
\end{abstract}

\section{Introduction}

It has been shown \cite{Belavin}, \cite{Vega},
 \cite{Bogomol'nyi} that in some
models solutions can be obtained by considering
   the first order differential equations, which are called Bogomol'nyi
   equations,     instead of  more complicated
   Euler-Lagrange  equations.
   The traditional method of obtaining such equations is based on
    rewriting an expression for the energy of a field configuration,
    in such a way, that  there is a lower bound on it, which has topological
    nature.
    Field configurations which saturate this bound
    satisfy  Euler-Lagrange equations as well as  Bogomol'nyi
    equations.

Another way to obtain such equations has also been  pointed out
  \cite{Witten}, \cite{Hlousek}.
 This method is connected with a N=2 supersymmetric extension
of an investigated model, and Bogomol'nyi equations arise naturally during
 detailed analysis of the algebra of supercharges.
 It has been shown that
in this case the energy of field configuration
is bounded below by the central
charge of the supersymmetric algebra.
   This method is more powerful than the previous one. As a result
  of this approach, we know \cite{Witten} that a Bogomol'nyi bound on
  the energy is valid not only classically, but also quantum mechanically.
 Another interesting fact,
indicated by this method, is that topologically
  non-trival field configurations of a N=1 supersymmetric
  theory must satisfy   the Bogomol'nyi bound. This statement is
  based on the
 existence of the N=2 supersymmetric extension of theory, which
 is a N=1 supersymmetric and possesses a  topologically conserved
 current \cite{Spector}.
 This method has  been successfully applied to many models.

  As an example let us consider
 the Abelian Higgs model, which was studied in \cite{Edelstein}. This model
 possesses vortex solution
 which has a topological charge (quantized magnetic flux).
It was shown that the central charge of the N=2 version of
  this model is in fact its topological charge. Furthermore, the special
  relation between coupling constants in this model, which
   is indispensable for the
  existence of Bogomol'nyi equations, appear as the necessary
  condition for the existence of its  N=2 supersymmetric extension.

There have  been  considerable interest in Chern-Simons systems
  \cite{Dunne}.
 These systems typically possess topological charge, therefore they are good
 candidates for investigations by the supersymmetric method.
 The Chern-Simons model without the Maxwell term but with a special
 sixth-order Higgs potential has been studied in this way
   \cite{Weinberg}. It was found that requirement
 of existence of the N=2 SUSY version of this model
 leads to the special form of the previously mentioned potential.
 When we want to consider a more general case, we add the Maxwell
 term to the action.
 It was shown \cite{Lee} that when we do so we must  also add the
 kinetic term of a neutral scalar field to the action and considerably
 change the potential. This model, in fact, contains two previously
 mentioned models. The first one is obtained by putting
 coupling constant, which stays next to the Chern-Simons term, equal to
  zero.
 The second one is
 obtained by making suitable limit of coupling constants \cite{Lee}.
 Our aim is to study the Maxwell Chern-Simons model by using
 the supersymmetric method, and  find its  Bogomol'nyi equations.

It's worth to notice that also the Maxwell Chern-Simons theory with an
additional magnetic moment interaction was studied \cite{Navratil},
\cite{Cuhna}. In the first paper  Bogomol'nyi equations were found
 by means of
 the supersymmetric method. Nevertheless,
 the results from this paper cannot be compared with ours.
 In the second one the N=2 supersymetric extension
 was found via a dimensional reduction.
 This method
 is significantly different from the one used in our paper, and
 it is instructive to compare these two approaches.

 The plan of this paper is as follows: we start our considerations
 from the Abelian Higgs model with the Chern-Simons term. Then
 we construct the N=1 supersymmetric version of this model. After that
 we indicate the difficulties connected with construction of the N=2
 supersymmetric action, and we show how they can be understood
 and  avoided.
 This leads to the correct form of the Maxwell Chern-Simons
 action. Next we find the Noether current, and construct appropriate
 real spinorial supercharges. Finally, we show how Bogomol'nyi equations
 arise  from their algebra and explicitly find these equations .

\section{Conventions}

Our conventions are as follows.
 We use a metric with the signature $(+,-,-)$, the covariant derivative is
 defined as: $D_{\mu}=\partial_{\mu}-i e A_{\mu}$.

 We take Dirac matrices $(\gamma^{\mu})_{\alpha}^{\; \; \beta}$ to be
\begin{eqnarray}
\gamma^{0}=\left( \begin{array}{cc} 0 & -i \\ i & 0
\end{array} \right), \ \ \ \ \ \gamma^{1}=\left( \begin{array}{cc} 0 & i \\ i & 0
\end{array} \right),  \ \ \ \ \ \gamma^{2}=\left( \begin{array}{cc} i & 0 \\ 0 & -i
\end{array} \right).
\end{eqnarray}
They obey the following equation
\begin{eqnarray}
\gamma^{\mu} \gamma^{\nu}= g^{\mu\nu}+ i
\epsilon^{\mu \nu \lambda}\gamma_{\lambda}.
\end{eqnarray}

Superspace conventions are the same as those in \cite{Gates}, and are briefly
 listed below for the reader's convenience.
Spinor indices are lowered and raised
 by the second-rank antisymmetric symbol
$C_{\alpha \beta}$ in the following way: $\psi^{\alpha}=C^{\alpha \beta}
\psi_{\beta},\psi_{\alpha}=\psi^{\beta}C_{ \beta \alpha}$;
$C_{\alpha \beta}$ has the form:
\begin{eqnarray}(C_{\alpha \beta})=(-C_{\beta \alpha})=
\left( \begin{array}{cc} 0 & -i \\ i & 0
\end{array} \right)=(-C^{\alpha \beta}).
\end{eqnarray}
A scalar superfield $\Phi=(\phi,\psi,F)$ is defined as
\begin{eqnarray}
 \Phi(x^{\mu},\theta^{\alpha})=\phi(x) + \theta^{\alpha} \psi_{\alpha}(x)
   - \theta^{2} F(x),
\end{eqnarray}
where $\theta^{\alpha}$ is a real spinor, $\theta^2=\frac{1}{2}\theta^{\alpha}
\theta_{\alpha}$,  and $\alpha=0,1$.
\\ A vector superfield $V^{\alpha}=(A_{\mu},\rho^{\alpha})$ in the
 Wess-Zumino
gauge reads
\begin{eqnarray}
  V^{\alpha}(x^{\mu},\theta^{\alpha})=i\theta^{\beta}
  (\gamma^{\nu})_{\beta}^{\; \; \alpha} A_{\nu}(x)-\theta^{2}
   2 \rho^{\alpha}(x).
  \end{eqnarray}
The supercovariant derivative is $D_{\alpha}=
  \frac{\partial} {\partial \theta^{\alpha}}  +  i \theta^{\beta}
  (\gamma^{\mu})_{\alpha \beta} \partial_{\mu}$, and
  the gauge covariant supercovariant derivative is
  $\nabla_{\alpha}=D_{\alpha} - i e V_{\alpha}$.

\section{The model}

It was shown \cite{Lee} that there are Bogomol'nyi equations in
 the  model defined by the action
\begin{eqnarray}
\label{1}
  {\cal S} & = &  \int d^3x \left[-\frac{1}{4} F^{\mu \nu} F_{\mu \nu}
  +\kappa \varepsilon^{\mu \nu \sigma} \partial_{\mu} A_{\nu} A_{\sigma}
  +\frac{1}{2} (D_{\mu} \phi)^{*} (D^{\mu} \phi )\right. \nonumber \\& &
  \left. +\frac{1}{2} \partial_{\mu} N \partial^{\mu} N
  -\frac{e^{2}}{2} N^{2} |\phi|^{2}
  -\frac{e^2}{8}\left(-\frac{4 N \kappa}{e} + |\phi|^{2}-\phi_{0}^{2}\right)^2
   \right],
  \end{eqnarray}
where $\phi$ is a complex scalar field, $N$ is a neutral real
  scalar field and
 $A_{\mu}$ is a gauge field.

We want to stress the fact that there are no Bogomol'nyi equations in
 the Abelian
Higgs model, which was studied in
 \cite{Edelstein},  with the Chern-Simons term. The action of
 this model can be written as
\begin{eqnarray}
\label{2}
  {\cal S}' & = &  \int d^3x \left[-\frac{1}{4} F^{\mu \nu} F_{\mu \nu}
  +\kappa \varepsilon^{\mu \nu \sigma} \partial_{\mu} A_{\nu} A_{\sigma}
   +\frac{1}{2} (D_{\mu} \phi)^{*} (D^{\mu} \phi )
      -\lambda (|\phi|^{2}-\phi_{0}^{2})^2 \right].
\end{eqnarray}
Our aim is to show, using supersymmetric formalism, that
 in order to obtain such equations we have to modify
 the action ({\ref{2}}) to the form of the action ({\ref{1}}).
  Consequently, we start our calculations from
 the action ({\ref{2}}) and we are looking for its supersymmetric
 version.

\section{ N=1 and N=2 extensions}

To obtain Bogomol'nyi equations we must  find a N=2 supersymmetric extension
of our model. The connection between Bogomol'nyi equations and
 the supersymmetric form of the investigated model was
  explained
  \cite{Hlousek}. We will discuss it in the next section.

In this section we start
    our considerations from a N=1
  supersymmetric extension of ({\ref{2}}).
We construct the appropriate  action    from
 the complex scalar superfield $\Phi=(\phi,\psi,F)$,
  the real scalar superfield $\Omega=(N,\chi,D)$,
 and the vector superfield $V^{\alpha}=(A_{\mu},\rho^{\alpha})$.
 The N=1 version of ({\ref{2}}) reads
\begin{eqnarray}
\label{3}
  {\cal S}'_{N=1} & = & \int d^3x d^{2} \theta\ \left[ -\frac{1}{4} (\nabla^{\alpha}
  \Phi)^{*}(\nabla_{\alpha} \Phi)
-\frac{1}{4} (D^{\alpha} \Omega)^{*}(D_{\alpha} \Omega)
-\frac{\kappa}{4} V^{\alpha} D_{\beta} D_{\alpha} V^{\beta}
\right. \nonumber \\& & \left.
+\frac{1}{16} (D_{\beta} D^{\alpha} V^{\beta})
 (D_{\gamma} D_{\alpha}V^{\gamma})
+(2 \lambda)^{\frac{1}{2}} \phi_{0}^2 \Omega
- (2 \lambda)^{\frac{1}{2}} \Phi^{*} \Phi \Omega \right].
\end{eqnarray}
In terms of the components of the superfields it takes the form
\begin{eqnarray}
\label{100}
  {\cal S}'_{N=1} & = & \int d^3x \left[
  -\frac{1}{4} F^{\mu \nu} F_{\mu \nu}
  +\kappa \varepsilon^{\mu \nu \sigma} \partial_{\mu} A_{\nu} A_{\sigma}
  +\frac{1}{2} (D_{\mu} \phi)^{*} (D^{\mu} \phi )
  +\frac{1}{2} \partial_{\mu} N \partial^{\mu} N \right. \nonumber  \\ & &
   -\lambda ( |\phi|^{2}-\phi_{0}^{2})^{2}
  -4 \lambda N^2 |\phi|^{2}
  +\frac{i}{2} \bar{\psi} \slash \!\!\!\! D \psi
  +\frac{i}{2} \bar{\rho} \slash \!\!\! \partial \rho
  +\frac{i}{2} \bar{\chi} \slash \!\!\! \partial \chi \nonumber  \\ & &
  \left.
  -(2 \lambda)^{\frac{1}{2}} \bar{\psi} \psi N
  +\frac{i e}{2} (\bar{\psi} \rho \phi - \bar{\rho} \psi \phi^{*})
  -(2 \lambda)^{\frac{1}{2}}(\bar{\chi} \psi \phi^{*}+\bar{\psi} \chi \phi)
   +\kappa \bar{\rho} \rho \right].
\end{eqnarray}
The non-propagating fields F and D were eliminated by means of their
Euler-Lagrange equations of motion.
 The action  ${\cal S}'_{N=1}$ is invariant under the
following N=1 transformations
\begin{eqnarray}
\label{4}
    \delta \psi_{\alpha}
   =-2 (2 \lambda)^{\frac{1}{2}} N \phi \eta_{\alpha}
   + i \eta^{\beta}(\gamma^{\mu})_{\alpha \beta} D_{\mu} \phi, \ \ \
   \delta \phi=\bar{\eta} \psi, \nonumber \\
   \delta \chi_{\alpha}=-2 (2 \lambda)^{\frac{1}{2}}
   (|\phi|^{2}-\phi_{0}^{2})\eta_{\alpha}+i \eta^{\beta} (\gamma^{\mu})_{\alpha \beta}
   \partial_{\mu} N, \ \ \ \delta N=\frac{1}{2}(\bar{\eta}
    \chi+\bar{\chi} \eta), \nonumber \\
   \delta \rho^{\alpha}=\frac{i}{2} \varepsilon^{\mu \nu \lambda} F_{\mu \nu}
   (\gamma_{\lambda})^{\alpha \beta} \eta_{\beta}, \ \ \
   \delta A^{\mu}=\frac{i}{2}( \bar{\eta} \gamma^{\mu} \rho - \bar{\rho}
   \gamma^{\mu} \eta),
\end{eqnarray}
where $\eta^{\alpha}$ is a real infinitesimal spinor.
\\ Evidently, when we put
  all fermion fields, as well as the
field
  N, equal to zero, the action ({\ref{100}}) will
   have the same form as the action ({\ref{2}}).
Therefore, the action ({\ref{3}}) is in fact the N=1 extension
 of ({\ref{2}}).

 To find the N=2 extension of ({\ref{100}}) we require its
invariance
 under transformations ({\ref{4}}) with an infinitesimal complex
 spinor  $\xi^{\alpha}$ instead of the real $\eta^{\alpha}$.
At this point, it is useful to change notation. We introduce,
  following \cite{Edelstein}, the spinor field $\Sigma$
\begin{eqnarray}
\Sigma=\chi - i \rho.
\end{eqnarray}
The invariance under the N=2 transformations can be achived by rewritting
 the
action ({\ref{100}})
  in the terms of
   $\Sigma$, $\psi$, $\phi$, N, $A_{\mu}$, and demanding
its invariance under transformations
\begin{eqnarray}
\label{5}
\Sigma \longrightarrow e^{-i\beta} \Sigma, \ \ \
\psi \longrightarrow e^{-i\beta} \psi,
\end{eqnarray}
where $\beta$ is defined as follows: $\xi^{\alpha}= e^{i \beta}
\eta^{\alpha}$. Obviously, this requirement is
equivalent to the previous one.
\\ To apply this method we  rearranged the action ({\ref{100}})
to the form
\begin{eqnarray}
\label{6}
  {\cal S}'_{N=1} & = &  \int d^3x \left[-\frac{1}{4} F^{\mu \nu} F_{\mu \nu}
  +\kappa \varepsilon^{\mu \nu \sigma} \partial_{\mu} A_{\nu} A_{\sigma}
  +\frac{1}{2} (D_{\mu} \phi)^{*} (D^{\mu} \phi ) \right. \nonumber \\& &
  +\frac{1}{2} \partial_{\mu} N \partial^{\mu} N
  -\lambda(|\phi|^{2}-\phi_{0}^{2})^2
  -4 \lambda N^{2} |\phi|^{2}
    +\frac{i}{2} \bar{\psi} \slash \!\!\!\! D \psi \nonumber \\ & &
  +\frac{i}{2} \bar{\Sigma} \slash \!\!\!
   \partial \Sigma
  -(2\lambda)^{\frac{1}{2}} \bar{\psi} \psi N
  -\left(\frac{e}{4} + \frac{(2\lambda)^{\frac{1}{2}}}{2}\right)
  (\bar{\psi} \Sigma \phi + \bar{\Sigma} \psi \phi^{*}) \nonumber \\ & &
   \left.
  +\left(\frac{e}{4} - \frac{(2\lambda)^{\frac{1}{2}}}{2}\right)
  (\bar{\psi} \bar{\Sigma} \phi + \Sigma \psi \phi^{*})
  +\frac{\kappa}{2} \bar{\Sigma} \Sigma
  +\frac{\kappa}{4} (\bar{\Sigma}\bar{\Sigma}-\Sigma \Sigma) \right].
\end{eqnarray}
As a consequence of the term $\frac{\kappa}{4}
(\bar{\Sigma}\bar{\Sigma}-\Sigma
\Sigma)$, this action is not invariant under transformations ({\ref{5}})
  even if we assume that
\begin{eqnarray}
\label{8}
\lambda=\frac{e^{2}}{8}.
\end{eqnarray}
This relation is exactly the same as that in \cite{Edelstein}.
 To obtain the N=2 SUSY version of ({\ref{2}}),
  we add to the action ({\ref{3}})
  the following term
\begin{eqnarray}
\label{7}
\int d^3x d^2 \theta  \kappa \Omega \Omega =
\int d^3x [ 2 \kappa N D + \kappa \chi \chi ]=
\int d^3x [2 \kappa N D +\frac{\kappa}{2} \bar{\Sigma} \Sigma
-\frac{\kappa}{4} (\bar{\Sigma}\bar{\Sigma}-\Sigma \Sigma) ].
\end{eqnarray}
One sees that the term ({\ref{7}}) cancel the last term of
  ({\ref{6}}), but it contains a field D. As a result, this
  addition leads to the
  modification of the Higgs term in the action.
The action, constructed as a sum of ({\ref{3}}) and ({\ref{7}}), is
  invariant
  under the N=2 supersymmetric transformations if we impose condition
   ({\ref{8}}) on $\lambda$ and $e$, and can be written as
\begin{eqnarray}
\label{9}
  {\cal S}_{N=2} & = &  \int d^3x \left[-\frac{1}{4} F^{\mu \nu} F_{\mu \nu}
  +\kappa \varepsilon^{\mu \nu \sigma} \partial_{\mu} A_{\nu} A_{\sigma}
  +\frac{1}{2} (D_{\mu} \phi)^{*} (D^{\mu} \phi )\right. \nonumber \\& &
  +\frac{1}{2} \partial_{\mu} N \partial^{\mu} N
  -\frac{e^{2}}{2} N^{2} |\phi|^{2}
  -\frac{e^2}{8}\left(-\frac{4 N \kappa}{e} + |\phi|^{2}-\phi_{0}^{2}\right)^2
  \nonumber \\ & & \left.
  +\frac{i}{2} \bar{\psi} \slash \!\!\!\! D \psi
  -\frac{e}{2} \bar{\psi} \psi N
  + \kappa \bar{\Sigma} \Sigma
  +\frac{i}{2} \bar{\Sigma} \slash \!\!\!
   \partial \Sigma
   - \frac{e}{2} ( \bar{\psi} \Sigma \phi
  + \bar{\Sigma} \psi \phi^{*}) \right].
\end{eqnarray}
The N=2 supersymmetric transformations read
\begin{eqnarray}
\label{10}
\delta \psi_{\alpha}
   =-e N \phi \xi_{\alpha}
   + i \xi^{\beta}(\gamma^{\mu})_{\alpha \beta} D_{\mu} \phi, \ \
   \delta \phi=\bar{\xi} \psi, \ \
   \delta A^{\mu}=-\frac{1}{2}( \bar{\xi} \gamma^{\mu} \Sigma + \xi
   \gamma^{\mu} \bar{\Sigma}),
   \nonumber \\
   \delta \Sigma_{\alpha}=(2 N \kappa - \frac{e}{2}
   (|\phi|^{2}-\phi_{0}^{2}))\xi_{\alpha}
   +(\frac{1}{2} \varepsilon^{\mu \nu \lambda} F_{\mu \nu}(\gamma_{\lambda})
   _{\alpha}
   ^{\; \; \beta} -i (\gamma^{\mu})_{\alpha}^{\; \; \beta} \partial_{\mu} N)
   \xi_{\beta}, 
   \nonumber \\
   \delta N=\frac{1}{2}(\bar{\xi}
    \Sigma+\bar{\Sigma} \xi).
\end{eqnarray}
 Now, if we put all fermion fields equal to  zero, we can see that the
requirement that the action ({\ref{3}}) must be
 invariant under the N=2 the transformations leads to  the
 action that is the supersymmetric extension of the initial
 one ({\ref{1}}).

\section{Bogomol'nyi equations}

Following Hlousek and Spector \cite{Hlousek} we concisely explain
  how Bogomol'nyi equations arise from the algebra of supersymmetric
  charges. Due to the Haag-$\L$opusza$\grave{n}$ski-Sohnius theorem \cite{Haag},
  there are two real spinorial
  supercharges $q_{\alpha}^{L}$, where $L=0,1$ is an internal index,
  in our  N=2 supersymmetric
  theory.
  These supercharges are obtained from the Noether conserved current.
  They obey the algebra
\begin{eqnarray}
\label{11}
\{q^{L}_{\alpha},q^{M}_{\beta}\}=2 \delta ^{L M} (\gamma^{\mu})_{\alpha
\beta}P_{\mu} + T C^{L M} C_{\alpha \beta},
\end{eqnarray}
where if we denote energy-momentum tensor by $T_{\mu \nu}$,
 we have $P_{\mu}=
\int d^2x T_{0 \mu}$. $T$ is the central charge.

The Noether current $(J^{\mu})^{\alpha}$ was found by considering variation of
the action ({\ref{9}}) under transformations ({\ref{10}}) with space-time
dependent spinorial parameter $\xi^{\alpha}$
\begin{eqnarray}
\delta {\cal S}_{N=2}= \int d^3x [(J^{\mu})^{\alpha}
\partial_{\mu} \xi_{\alpha} + h.c.  ].
\end{eqnarray}
In the present case we have
\begin{eqnarray}
\label{20}
(J^{\mu})^{\alpha}& = &
    \frac{1}{2} \bar{\Sigma}^{\beta} (\gamma_{\nu})_{\beta}^{\; \; \alpha}
     F^{\mu \nu}
    +\frac{i}{2} \varepsilon^{\mu \nu \lambda} \bar{\Sigma}^{\beta}
    (\gamma_{\lambda})_{\beta}^{\; \; \alpha} \partial_{\nu} N
    +\frac{i}{4} \varepsilon^{\mu \nu \rho} F_{\nu \rho}
     \bar{\Sigma}^{\alpha} \nonumber \\ & &
     + i \kappa \bar{\Sigma}^{\beta}(\gamma^{\mu})_{\beta}^{\; \; \alpha} N
     +\frac{1}{2} D^{\mu} \phi \bar{\psi}^{\alpha}
     + \frac{1}{2} \partial^{\mu} N \bar{\Sigma}^{\alpha}
     -\frac{i e}{2} \bar{\psi}^{\beta}(\gamma^{\mu})_{\beta}^{\; \; \alpha}
     \phi N \nonumber \\ & &
     -\frac{i e}{4} \bar{\Sigma}^{\beta}(\gamma^{\mu})_{\beta}^
     {\; \; \alpha} (|\phi|^{2}-\phi_{0}^{2})
     +\frac{i}{2} \varepsilon^{\mu \nu \lambda} \bar{\psi}^{\beta}
     (\gamma_{\lambda})_{\beta}^{\; \; \alpha} D_{\nu} \phi,
\end{eqnarray}
and $\partial_{\mu}(J^{\mu})^{\alpha}=0$.

 The supercharges are defined by the
relations
\begin{eqnarray}
\label{21}
q_{\alpha}^{1}= \int d^2x [ (J^{0})_{\alpha} + h.c.],
\ \ \ q_{\alpha}^{2}=
-i \int d^2x [ (J^{0})_{\alpha} - h.c. ].
\end{eqnarray}
In order
 to check the relation ({\ref{11}}),
 we must impose canonical (anti)commutation relations
 on our fields.
If we simplify  calculations by putting all fermion fields zero after
computing the anticommutator ({\ref{11}}), we only need the following
canonical    anticommutation relation for the field $\Sigma$
\begin{eqnarray}
\label{101}
\{ \Sigma^{\beta}(\vec{x}), \frac{i}{2}
(\gamma^{0})_ {\alpha}^{\; \; \sigma}\bar{\Sigma}_{\sigma}(\vec{y}) \}
= i \delta^{2}(\vec{x}-\vec{y}) \delta_{\alpha}^{\; \; \beta},
\end{eqnarray}
and the same relation for the field $\psi$. We
 don't use a special symbol for operators.
 It should be noticed that the relation
({\ref{101}}) is valid when $\Sigma$ is an operator, so this equation must
 be understood as the operator equation.
 If not stated
 otherwise, the expressions below
 are the operator
 equations, except the case when they contain the expectation value, which
 is  denoted by $\langle \; \rangle$.
\\ After lenghty  but straightforward calculations, one obtains
\begin{eqnarray}
\label{12}
 \langle P_{0} \rangle & = & \int d^2x \left[ \frac{1}{4} (F_{i j})^{2}
           +\frac{1}{2}(F_{i 0})^{2}
           +\frac{1}{2} |D_{0}\phi|^{2}
           +\frac{1}{2} |D_{i}\phi|^{2}
           +\frac{1}{2} (\partial_{0} N)^{2}
           +\frac{1}{2} (\partial_{i} N)^{2} \right. \nonumber \\ & &
           \left.
           +\frac{e^{2}}{2} |\phi|^{2} N^{2}
           +\frac{e^{2}}{8} \left(-\frac{4 N \kappa}{e}
           + |\phi|^{2}-\phi_{0}^{2}\right)^2  \right],
\end{eqnarray}
\begin{eqnarray}
\label{13}
\langle P_{i} \rangle = \int d^2x \left[ -F_{0}^{\;\; k} F_{i k} + \partial_{0} N \partial_{i} N
      +\frac{1}{2} D_{0} \phi (D_{i} \phi)^{*}+\frac{1}{2}
      (D_{0} \phi)^{*} D_{i} \phi \right],
\end{eqnarray}
\begin{eqnarray}
\label{14}
 T  =\int d^2x \varepsilon^{i j}\partial_{j} ( e \phi_{0}^{2} A_i -
            i \phi^{*} D_{i}\phi  )
   =  e \phi_{0}^{2} \Phi,
\end{eqnarray}
where indices $i,j,k=1,2$. $\Phi=- \int d^2x F_{1 2}$, the magnetic flux,
  is   the topological charge of the Maxwell
 Chern-Simons theory \cite{Lee}.
 What's more,
since the central charge $T$ is a scalar, expression ({\ref{14}}) contains
classical
 fields.
 To attain the exact form of the central charge
 $T$  Euler-Lagrange equations
of motion for the field N have been used. We have also assumed that
 $D_{i}\phi$
tends to zero at infinity.

Now, we are ready to find Bogomol'nyi equations. Let us  introduce
 $Q_{1}$ and $Q_{2}$ by
\begin{eqnarray}
\label{16}
 Q_{1}=\frac{1}{2}(q_{1}^{1}+i q_{1}^{2}), \ \ \
 Q_{2}=\frac{1}{2}(q_{2}^{2}-i q^{1}_{2}).
\end{eqnarray}
Hence
\begin{eqnarray}
\label{17}
\{ Q_{1} \pm Q_{2},
 Q_{1}^{\dagger} \pm Q_{2}^{\dagger} \} = 2\left(P_{0}\pm \frac{T}{2}\right),
\end{eqnarray}
where a unit operator next to $T$ is not written; the lower (upper)  sign
corresponds to a positive (negative)
 value of  $T$.
\\ Taking the expectation value of ({\ref{17}), one can conclude that there is
  a Bogomol'nyi bound
\begin{eqnarray}
\langle P_{0}\rangle\geq \frac{| T  |}{2}.
\end{eqnarray}
Moreover, this bound is saturated when
\begin{eqnarray}
\label{22}
(Q_{1} \pm Q_{2})|B\rangle = 0.
\end{eqnarray}
Using the relations ({\ref{20}), ({\ref{21}), ({\ref{16}), ({\ref{22}), one
finds
\begin{eqnarray}
 F_{i 0} \mp \partial_{i} N =0, \nonumber \\
 F_{1 2} \mp 2 ( \kappa N - \frac{e}{4}(|\phi|^{2}-\phi_{0}^{2}))=0,
 \nonumber \\
 (D_{1} \pm i D_{2} )\phi=0, \nonumber \\
 D_{0}\phi \pm i e \phi N =0, \nonumber \\
 \partial_{0} N =0,
\end{eqnarray}
where $\phi, N, A_{\mu}$ are classical fields.
 These equations are precisely the Bogomol'nyi equations that we
  were looking for. They are of course the same as those in \cite{Lee}.
  We want to emphasize that during these calculations we didn't
  choose any particular gauge choice for the field $A_{\mu}$, and we didn't
  assume that our fields are time-independent,
  as it was done in \cite{Edelstein}.

To summarize, the supersymmetric method of finding Bogomol'nyi
 equations next time turned out to be a useful tool.
 We saw that a special form of the potential term
  and an absolute necessity of the additional real neutral scalar field,
  in considered  model,
 is an artifact of the existence of its N=2 supersymmetric extension.
 We also checked that the topological charge of the considered model
 is in fact the central charge of its supersymetric extension.

 I would like to thank Leszek Hadasz for
 helpful discussions and reading the manuscript.


\begin{thebibliography}{99}

\bibitem{Belavin}
A.A. Belavin, A.M. Polyakov, A.S. Schwartz and Yu.S. Tyupkin,
   Phys. Lett. B {\bf 59} (1975) 85.

\bibitem{Vega}
H. J. de Vega and F.A. Schaposnik,
 Phys. Rev. D {\bf 14} (1976) 1100.

\bibitem{Bogomol'nyi}
E.B. Bogomol'nyi Yad. Fiz 24 (1976) 861.

\bibitem{Witten}
D. Olive and E. Witten, Phys. Lett. B {\bf 78} (1978) 97.

\bibitem{Hlousek}
Z. Hlousek and D. Spector, Nucl. Phys. B {\bf 397} (1993) 173.

\bibitem{Spector}
Z. Hlousek and D. Spector, Nucl. Phys. B {\bf 370} (1992) 143.

\bibitem{Edelstein}
J. Edelstein, C. N$\acute{u}\tilde{n}$ez and F. Schaposnik, Phys. Lett. B {\bf
329}
 (1994) 39.

\bibitem{Dunne}
G. V. Dunne,  hep-th/9902115.

\bibitem{Weinberg}
C. Lee, K. Lee and E.J. Weinberg, Phys. Lett. B {\bf 243} (1990) 105.

\bibitem{Lee}
C. Lee, K. Lee and  H. Min,
  Phys. Lett. B {\bf 252} (1990) 79.

\bibitem{Navratil}
P. Navr$\acute{a}$til, Phys. Lett. B {\bf 365} (1996) 119.

\bibitem{Cuhna}
H.R. Christiansen, M.S. Cunha, J.A. Helay$\ddot{e}$l-Neto, L.R.U. Manssur
 and A.L.M.A Nogueira,
 Int. J. of Mod. Phys. A {\bf 14} (1999) 147.


\bibitem{Gates}
S.J. Gates, M.T. Grisaru, M. Ro$\hat{c}$ek and W.Siegel, Superspace,
 or one thousand and one lessons in supersymmetry
(Benjamin/Cummings, Menlo Park, 1983).

\bibitem{Haag}
R. Haag, J.T. $\L$opusza$\acute{n}$ski and M. Sohnius, Nucl. Phys. B {\bf 88}
(1975) 257.

\end{thebibliography}
\end{document}